\documentclass[twocolumn,showpacs,amsmath,amssymb,aps,pra,floatfix]{revtex4}
\usepackage{bm}
\usepackage{amsfonts}
\usepackage{graphicx}
\usepackage{graphicx}

\begin{document}

\title{Infrared self-consistent solutions of bispinor QED$_3$}
\author{Tomasz Rado\.zycki}
\email{t.radozycki@uksw.edu.pl} \affiliation{Faculty of Mathematics and Natural Sciences, College of Sciences,
Cardinal Stefan Wyszy\'nski University, W\'oycickiego 1/3, 01-938 Warsaw, Poland}

\begin{abstract}
Quantum electrodynamics in three dimensions in the bispinor formulation is considered. It is shown that the Dyson-Schwinger equations for fermion and boson propagators may be self-consistently solved in the infrared domain if on uses the Salam's vertex function. The parameters defining the behavior of the propagators are found numerically for different values of coupling constant and gauge parameter. For weak coupling the approximated analytical solutions are obtained. The renormalized gauge boson propagator (transverse part) is shown in the infrared domain to be practically gauge independent.
\end{abstract}
\maketitle

\section{Introduction}

QED in three space-time dimensions has become a testing laboratory for certain nonperturbative aspects of QFT like chiral symmetry breaking~\cite{pis,abcw,abkw,kye,koma,bas1,bas2,bas3,bas4,bas5}, bound states~\cite{ab,matna,hosh} or confinement~\cite{bas3,gm,br,bpr,maris,sauli} and for various approximation schemes. Due to the dimensionality of the coupling constant (as $\sqrt{\mathrm{mass}}$) it is a superrenormalizable theory so it is free of infinite renormalization ambiguities typical for four-dimensional theory.  

It can be formulated in two inequivalent versions: with two- and four-component fermions~\cite{arm,bas6}. The properties of the theory are different in these two cases. To the investigation of both versions much attention has been payed over the last twenty years. The work has been especially concentrated on nonperturbative solutions of Dyson-Schwinger (DS) equations with different approximations incorporated in the theory as quenched approximation, rainbow approximation, $1/N$ expansion and various models of vertex function ~\cite{pis,abcw,abkw,kye,bas1,bas5}. Particularly often the  multi-flavor theory with the limit $N\rightarrow\infty$ has been used, since it avoids infrared problems, which usually become troublesome in lower number of dimensions. The other important point in this analysis has been the unpleasant gauge dependence of the nonperturbative results which constitute the common problem of the approximated studies based on DS equations~\cite{bpr,bas1,bara2,barasm,bas4,bas5}.

As it is well known, DS equations constitute an infinite set of relations involving Green's functions which form the `inverted ladder' type structure: the $n$-point functions depend on $n+1$-point ones and so on up to infinity. This set cannot be solved without the truncation of such a hierarchy. But even such truncation, which turns the infinite set of equations into only a couple of them, leads to the system which is far from being trivial and requires further simplifications. Usually this truncation is accomplished by performing certain assumption for the vertex function, which is chosen most often in the form satisfying Ward-Takahashi identity. Naturally this identity fixes only longitudinal part of the vertex leaving the transverse part a subject of further discussion and improvements~\cite{bc,dz,park,cp,br,bara}. 

 In our previous paper~\cite{tr3} we applied a method elaborated earlier in QED$_4$~\cite{tribb} to QED$_3$ with two-component fermions. It consists on the following five steps:
\begin{enumerate}
\item a certain infrared form of two basic propagators, $S(p)$ and $D^{\mu\nu}(k)$ (suggested by perturbative calculations or other methods) as dependent on a couple of unknown parameters is assumed,
\item the fermion propagator is represented in the spectral form with one known spectral density $\rho(M)$, which is in general possible in the infrared domain,
\item Salam's form of the vertex function~\cite{sal2}, together with $S(p)$ and $D^{\mu\nu}(k)$, is substituted into the first two of the set of DS equations,
\item from these two equations the set of self-consistent equations for parameters is derived,
\item the obtained equations are solved numerically or analytically.
\end{enumerate}
This method proved to be relatively effective both in QED$_4$, where all parameters were correctly found without the necessity of infinite renormalization, and in spinor version of QED$_3$, where obtained results stay in general agreement with other works. In the present paper we would like to extend its application to four-component QED$_3$, without Chern-Simons term, i.e. when gauge bosons (`photons') remain massless in spite of interaction. 

The common effect of various simplifications of DS equations is the gauge dependence of the results (even of the physical observables). The sources of this undesirable behavior are the approximations made to the propagators and to the vertex. It seems therefore valuable to test various possible approaches with regard to that particular feature and much work has already been done in this direction (see the references above). This point lies in the scope of interest of the present work too.

This paper is organized as follows. In the next section we define the model itself and give the resulting set of DS equations. In section~\ref{sec:assump} the infrared Green's function in question are formulated up to several unknown parameters. In section~\ref{sec:solu} we substitute these Green's functions into DS equations and obtain the set of relations for the introduced parameters. In the last section we present analytical and numerical results and some conclusions.

\section{Formulation of the model.}
\label{sec:formo}

The model is defined through following Lagrangian density :
\begin{eqnarray} 
{\cal L}(x)=&&\!\!\!\!\!\overline{\Psi}(x)\left(i\gamma^{\mu}\partial_{\mu}-m_0 -
e_0\gamma^{\mu}A_{\mu}(x)\right)\Psi (x)\label{eq:lagr} \\
&&\!\!\!\!\!\!-
\frac{1}{4}F^{\mu\nu}(x)F_{\mu\nu}(x)-
\frac{\lambda}{2}\left(\partial_{\mu}A^{\mu}(x)\right)^2\; , 
\nonumber
\end{eqnarray} 
where $\lambda$ is the gauge parameter. The quantities $m_0$ and $e_0$ denote here the {\em bare} fermion mass and the {\em bare} coupling constant respectively. As mentioned in the Introduction, the latter for $D=2+1$ is a quantity with the dimensionality of $\sqrt{\mathrm{mass}}$. That means that in the quantum theory higher terms of perturbation expansions have better ultraviolet momentum dependence in loop integrations and the model is superrenormalizable.
 
As already told, in the present paper we deal with four-component fermion field, choosing the  following representation for gamma matrices used also in four-dimensional QED:
\begin{eqnarray} 
&&\gamma^0=\left(\begin{array}{ccc}\sigma_3 & & 0 \\ 0 & & -\sigma_3 
\end{array}\right)\; , \;\;\;\;\; 
\gamma^1=\left(\begin{array}{cc} i\sigma_1 & 0 \\ 0 & -i\sigma_1 
\end{array}\right)\; , \nonumber\\ 
&&\gamma^2=\left( 
\begin{array}{cc} \hspace*{1ex}i\sigma_2 & 0 \\ 0 & -i\sigma_2 \end{array}\right) \; .
\label{eq:gammas}
\end{eqnarray}

There are two other gamma matrices, which anticommute with all above, and which can serve for defining chiral transformations:
\begin{equation} 
\gamma^3=\left(\begin{array}{ccc}0 & & \openone \\ \openone & & 0 
\end{array}\right)\; , \;\;\;\;\; 
\gamma^5=\left(\begin{array}{cc} 0&  i\openone  \\ -i\openone & 0 
\end{array}\right)\; .
\label{eq:gamma35}
\end{equation}

The fermion mass term, that was chosen in~(\ref{eq:lagr}) in the form $m_0\overline{\Psi}\Psi$, breaks the chiral symmetry defined by any of the matrices~(\ref{eq:gamma35}). There is, however, the possibility of the other choice for the mass term~\cite{abkw}:
$m'_0\overline{\Psi}\tau \Psi$, where
\begin{equation}
\tau=\frac{i}{2}\left[\gamma^3,\gamma^5\right]=\left(\begin{array}{cc}\openone  &  0\\ 0 & -\openone   
\end{array}\right)\; .
\label{eq:tauma}
\end{equation}
This term does not break chiral symmetry, since matrix $\gamma^0\tau$ commutes with both $\gamma^3$ and $\gamma^5$, but it does violate parity symmetry (contrary to the choice made in~(\ref{eq:lagr})).

For matrices~(\ref{eq:gammas}) we have ordinary relations:
\begin{eqnarray}
\{\gamma^\mu, \gamma^\nu\}=g^{\mu\nu}\; ,\;\;  \mathrm{tr}\,\gamma^\mu=0\; ,\;\; \mathrm{Tr}\,\left[\gamma^\mu \gamma^\nu\right]=4g^{\mu\nu}\; ,\nonumber\\
\mathrm{Tr}\,\left[\gamma^\mu \gamma^\nu \gamma^\rho \gamma^\sigma\right]=4\left(g^{\mu\nu}g^{\rho\sigma}-g^{\mu\rho}g^{\nu\sigma}+g^{\mu\sigma}g^{\nu\rho}\right)\;,
\label{eq:toga}
\end{eqnarray} 
where we choose for the metric tensor:
\begin{eqnarray}
g^{00} &\!\! = &\!\! -g^{11}=-g^{22}=1\; .
\label{eq:mette}
\end{eqnarray}

The trace of the product of an odd number of gamma matrices equals zero as in four dimensions (this was not the case in the spinor representation, where it was proportional to the antisymmetric tensor $\varepsilon^{\mu\nu\rho}$). Additionally we have the identity:
\begin{equation}
\gamma^\mu \gamma^\nu \gamma_\mu =-\gamma^\nu\; .
\label{eq:ggg}
\end{equation}

\begin{figure}[h]
\centering
{\includegraphics[width=0.48\textwidth]{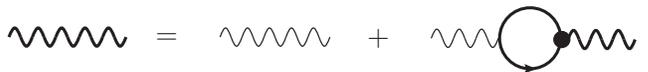}
\caption{Dyson-Schwinger equation for the gauge boson propagator $D^{\mu\nu}(k)$. Light lines represent free propagators and heavy ones dressed propagators. The full circle stands for the full fermion-boson vertex.} \label{fig:dspf}}
\end{figure}

Using the Lagrangian density~(\ref{eq:lagr}), one can derive in the standard way -- for instance through Feynman path integral -- the Dyson-Schwinger equations for propagators~\cite{iz}. For boson propagator we obtain the relation
\begin{eqnarray}
&&\!\!\!\!\!D^{\mu\nu}(k)=\frac{1}{k^2}\left(-g^{\mu\alpha}+
\frac{k^{\mu}k^{\alpha}}{k^2}-\frac{1}{\lambda}
\frac{k^{\mu}k^{\alpha}}{k^2}\right)\bigg[\delta_{\alpha}^{\nu}-ie^2_0\times\nonumber\\
&&\!\!\!\!\!\times\mathrm{Tr}\,\gamma_{\alpha}\int\frac{d^3p}{(2\pi)^3}S(p)\Gamma_{\beta}(p,
p-k)S(p-k)D^{\beta\nu}(k)\bigg],\nonumber\\
\label{eq:dysph}
\end{eqnarray}
which may be given the graphical form shown in Figure~\ref{fig:dspf}. The propagator $D^{\mu\nu}(k)$ on the right hand side is not integrated over three-momenta and therefore it can fully be represented through fermion functions (and the vertex).

The DS equation for the fermion propagator does not allow for such a separation, and has the form
\begin{eqnarray}
S(p)=&&\!\!\!\!\! 
\frac{1}{\not\! p - m_0}\bigg[1+ie^2_0\gamma^{\mu}\times\label{eq:dyse}\\
&&\!\!\!\!\!\times\int\frac{d^3k}{(2\pi)^3}S(p+k)\Gamma^{\nu}(p
+k, p)S(p)D_{\mu\nu}(k)\bigg]\; ,
\nonumber
\end{eqnarray}

\begin{figure}[h]
\centering
{\includegraphics[width=0.48\textwidth]{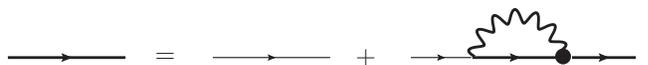}
\caption{Dyson-Schwinger equation for the fermion propagator. As in Figure~\ref{fig:dspf}, heavy lines stand for full functions and light for free ones.} \label{fig:dsef}}
\end{figure}
Its graphical representation is shown in Figure~\ref{fig:dsef}. While introducing the fermion self-energy $\Sigma(p)$, it can be rewritten in a simpler manner
\begin{equation}
(\not\! p -m_0)S(p)=1+\Sigma(p)S(p)\; ,
\label{eq:hs}
\end{equation}
where
\begin{eqnarray}
&&\!\!\!\!\!\Sigma(p)S(p)=\label{eq:sis}\\
&&=ie^2_0\gamma^{\mu}\int\frac{d^3k}{(2\pi)^3}S(p+k)
\Gamma^{\nu}(p+k, p)S(p)D_{\mu\nu}(k)\; .
\nonumber
\end{eqnarray}

Higher equations are not considered in our present approach.  The vertex function $\Gamma^\mu$ is not taken from DS equations, but postulated in the form proposed by Salam~\cite{sal2} and used afterwards in the so called `gauge technique'~\cite{del1,tz,dz,park}. The eventual extension of our method on higher Green's functions will be considered elsewhere.

\section{Infrared Green's functions}
\label{sec:assump}

The free propagator of massless vector boson has the standard form:
\begin{equation}
D^{(0)\mu\nu}(k) = \frac{1}{k^2}\left(-g^{\mu\nu}+
\frac{k^{\mu}k^{\nu}}{k^2}\right) -
\frac{1}{\lambda} \frac{k^{\mu}k^{\nu}}{(k^2)^2}\; .
\label{eq:dprfree}
\end{equation}

As is well known, the interaction with fermions does not change the longitudinal part of $D^{\mu\nu}(k)$. This is guaranteed by the following Ward-Takahashi (WT) identity
\begin{equation}
k_\mu D^{\mu\nu}(k)=k_\mu D^{(0)\mu\nu}(k)=-\frac{1}{\lambda}\,\frac{k^\nu}{k^2}\; .
\label{eq:wtid}
\end{equation}
Consequently the nonperturbative propagator may be then written as
\begin{equation}
D^{\mu\nu}(k) = \frac{Z_3}{d(k^2)}\left(-g^{\mu\nu}+
\frac{k^{\mu}k^{\nu}}{k^2}\right) -
\frac{1}{\lambda} \frac{k^{\mu}k^{\nu}}{(k^2)^2}\; ,
\label{eq:dprfull}
\end{equation}
with certain unknown function $d(k^2)$. 
In the infrared domain we assume this function to have a Taylor expansion starting from $k^2$, since for bispinor fermions the boson field remains massless~\cite{bpr} and we expect $d(0)=0$. This is not the case for massive two-component fermions, for which the topological boson mass is generated. 

Taking the first two terms of this expansion, we have
\begin{equation}
d(k^2)=k^2\left(1+\frac{k^2}{\kappa^2}\right)\; ,
\label{eq:dexp}
\end{equation}
where $\kappa^2$, together with the renormalization constant $Z_3$, will be determined from consistency conditions.
From the unitarity we expect the value of $Z_3$ to satisfy the restriction $0<Z_3\leq 1$~\cite{wei}. This expectation was actually confirmed in our previous works on spinor QED$_3$~\cite{tr3} and QED$_4$~\cite{tribb}. 

With the above assumptions the inverse of $D^{\mu\nu}$ may be written as
\begin{eqnarray}
D^{-1}&&\!\!\!\!\!\!\!  (k)^{\mu\nu}=\label{eq:dinv}\\
&& Z_3^{-1}\bigg(1+\frac{k^2}{\kappa^2}\bigg)(-k^2g^{\mu\nu}+k^{\mu}k^{\nu})-\lambda k^\mu k^\nu\; .
\nonumber
\end{eqnarray}
 
The free fermion propagator has the usual form:
\begin{equation}
S^{(0)}(p) = \frac{1}{\not\! p - m_0}\; .
\label{eq:sprfree}
\end{equation}
Due to the masslessness of the gauge boson and the absence of the mass gap for emission of soft photons, the pole at mass $m_0$ should, in the full propagator, turn into a branch point at $p^2=m^2$, where $m$ is a physical mass. This is expected also on the basis of general considerations on the analytical structure of fermion propagator~\cite{maris2, maris} and of confinement, which prohibits $S(p)$ from having a simple pole. The numerical results, performed in $1/N$ expansion with bare vertex and in euclidean space, suggesting the existence of {\em complex} singularities rather than {\em real} may as well constitute the effect of the coarse approximations made while solving DS equations: the approximations that are inevitable in any nonperturbative approach. Such singularities may be a signal of confinement, but are not prerequisite. In the Schwinger Model, massless electrodynamics in two space-time dimensions, which exhibits confinement of fermions, the {\em real} singularity in the infrared domain in the fermion propagator has been found (it has the form of ~$1/(-p^2)^{5/4}$)~\cite{stam,trpert}. 

Our infrared assumption for $S(p)$ is then:
\begin {equation}
S(p) = \frac{1}{(\not\! p - m) (1-p^2/m^2)^{\beta}}\; .
\label{eq:spr}
\end{equation}
We will see later that it will turn out to be self-consistent. 

The values of the exponent $\beta$ and of mass renormalization constant $\delta m=m-m_0$  will be established by the requirement of consistency. The presence of $\beta$ in denominator improves the ultraviolet  behavior of loop integrals (we assume that $0<\beta<1$, to be verified {\em a posteriori}). 

For our further purposes $S(p)$ has to be written in the spectral form
\begin{equation}
S(p) = \int dM\rho (M)\left(\frac{1}{\not\! p - m} -
\frac{1}{\not\! p - M}\right)\; ,
\label{eq:sp}
\end{equation}
with the one spectral density :
\begin{eqnarray}
\rho(M) =&&\!\!\!\!\frac{\sin(\pi\beta)}{\pi}\frac{1}{(M-m)
(M^2/m^2-1)^{\beta}}\nonumber\\
&&\!\! \times [\Theta (M-m) - \Theta (-M-m)]\;,
\label{eq:ro}
\end{eqnarray}
sufficient to define the infrared behavior of the propagator. $\Theta$ is here the Heaviside step function. The particular form of $\rho(M)$ was given in our previous works~\cite{tribb,tr3}.

Using $\rho(M)$, one can write the vertex function, which we need to put into DS equations~(\ref{eq:dysph})~and~(\ref{eq:dyse}). As mentioned in the Introduction we use the Salam's vertex~\cite{sal2} (with slight and obvious modification resulting from the form~(\ref{eq:sp}) of $S(p)$, which in our case contains two terms):
\begin{eqnarray}
S(p+k)&&\!\!\!\!\! \Gamma^{\mu}(p+k, p)S(p)\nonumber\\
&&\!\!\!\!\! = \int dM\rho (M)
\Bigg[ \frac{1}{\not\! p\; + \not\! k -m}\gamma^{\mu}\frac{1}
{\not\! p - m}\nonumber\\
&& - \frac{1}{\not\! p\;  + \not\! k -
M}\gamma^{\mu}\frac{1}{\not\! p - M}\Bigg] \; . 
\label{eq:verp}
\end{eqnarray}
It automatically guarantees the compliance with the 
WT identity:
\begin{equation}
k_\mu S(p+k)\Gamma^{\mu}(p+k, p)S(p)=S(p)-S(p+k)\; .
\label{eq:kve}
\end{equation}

\section{Self-consistent infrared equations}
\label{sec:solu}

\subsection{Gauge boson propagator}
\label{subsec:gbp}

Inserting~(\ref{eq:sp}) and~(\ref{eq:verp}) into the right hand side of~(\ref{eq:dysph}) we obtain

\begin{eqnarray}
&&\!\!\!\!\!D^{-1}(k)^{\mu\nu}=
-k^2g^{\mu\nu}+k^{\mu}k^{\nu}-\lambda k^{\mu}
k^{\nu }
+ie^2_0\mathrm{Tr}\,\gamma^{\mu}\times\nonumber\\ 
&&\!\!\times\int dM\rho(M)\int\frac{d^3p}{(2\pi)^3}\Big(\frac{1}
{\not\! p - m + i\varepsilon}\gamma^{\nu}\frac{1}{\not\! p\;  -
\not\! k - m + i\varepsilon} \nonumber\\
&&\!- \frac{1}{\not\! p - M +i
\varepsilon}\gamma^{\nu}\frac{1}{\not\! p\;  - \not\! k - M +
i\varepsilon}\Big)\; , \label{eq:invpp}
\end{eqnarray}
where we have rewritten this equation for inverse propagator, which is easier to handle, because in this case $D^{\mu\nu}$ decouples from other functions. 

The vacuum polarization tensor may be defined as the spectral integral  
\begin{equation}
\Pi^{\mu\nu}(k)=\int dM\rho(M)\left(\Pi_m^{\mu\nu}(k)-\Pi_M^{\mu\nu}(k)\right)\; .
\label{eq:pote}
\end{equation}
where $\Pi_m^{\mu\nu}(k)$ denotes the usual perturbative tensor:
\begin{eqnarray}
&&\!\!\!\!\!\Pi_m^{\mu\nu}(k)=\label{eq:pimn}\\
&&ie^2_0\mathrm{Tr}\,\gamma^{\mu}\int\frac{d^3p}{(2\pi)^3}\frac{1}
{\not\! p - m + i\varepsilon}\gamma^{\nu}\frac{1}{\not\! p\;  -
\not\! k - m + i\varepsilon}\; .
\nonumber
\end{eqnarray}

It may be evaluated in the standard way with the use of identities~(\ref{eq:toga}), and for instance by performing Wicks rotation and introducing Feynman parameters. Passing back to Minkowski space, we have
\begin{eqnarray}
&&\!\!\!\!\!\Pi_m^{\mu\nu}(k)=\label{eq:pimnf}\\
&&\!\!\!\frac{e_0^2}{\pi}(-k^2g^{\mu\nu}+k^{\mu}k^{\nu})\int\limits_0^1 dx\frac{x(1-x)}{(m^2-k^2x(1-x))^{1/2}}\; .
\nonumber
\end{eqnarray}

For $k^2<4m^2$ the $x$ integral is well defined. The transversality of $\Pi_m^{\mu\nu}(k)$ is a favorable consequence of using the vertex function in the form~(\ref{eq:verp}), satisfying the WT identity~(\ref{eq:kve}).
Performing the spectral integral over $M$ in~(\ref{eq:pote}) with the use of~(\ref{eq:ro}), similarly as it was done in~\cite{tr3}, we obtain:
\begin{eqnarray}
&&\!\!\!\!\!\Pi^{\mu\nu}(k)=\frac{e_0^2\Gamma(\beta+1/2)}{\pi^{3/2}\Gamma(\beta+1)m}\times\label{eq:pmn}\\
&&\times\left(-k^2g^{\mu\nu}+k^\mu k^\nu\right)\int\limits_0^1 dx\frac{x(1-x)}{(1-k^2/m^2\, x(1-x))^{\beta+1/2}}\; .
\nonumber
\end{eqnarray}

The integral over Feynman parameter $x$ leads to the hypergeometric (Gauss) function
\begin{eqnarray}
\int_0^1 dx&&\!\!\!\!\!\!\frac{x(1-x)}{(1-y\,x(1-x))^{\beta+1/2}} =\nonumber\\
&&\;\;\;\frac{1}{6}\:{}_2F_1(2,\beta+1/2;5/2;y/4)\; ,
\label{eq:in1}
\end{eqnarray}
and the DS equation~(\ref{eq:invpp}) may be given the form
\begin{eqnarray}
&&\!\!\!\!\!\!D^{-1}(k)^{\mu\nu}=-\lambda k^{\mu}k^{\nu}+
(-k^2g^{\mu\nu}+k^{\mu}k^{\nu})\times\label{eq:dysftot}\\
&&\!\!\!\!\!\times\left[1+\frac{e_0^2\Gamma(\beta+1/2)}{6\pi^{3/2}\Gamma(\beta+1) m}\:{}_2F_1(2,\beta+1/2;5/2;k^2/4m^2)\right]\; .
\nonumber
\end{eqnarray}

After the substitution of the expression~(\ref{eq:dinv}) for the left hand side and cancellation of the tensor structures, we are left with the scalar equation, for which we require the adjustment the first two terms of the Taylor expansion in $k^2$. In that way we get two equations for $Z_3$ and $\kappa^2$:
\begin{eqnarray}
Z_3^{-1} =&&\!\!\!\!\! 1+\frac{e_0^2\Gamma(\beta+1/2)}{6\pi^{3/2} m\Gamma(\beta+1)}\; ,\label{eq:eq1}\\
Z_3^{-1}\frac{1}{\kappa^2} =&&\!\!\!\!\! \frac{e_0^2\Gamma(\beta+3/2)}{30\pi^{3/2} m^3\Gamma(\beta+1)}\; ,
\label{eq:eq2}
\end{eqnarray}
where we expanded the Gauss function for small momentum according to the formula:
$$
\:{}_2F_1(a,b;c;z)\approx 1+\frac{ab}{c}\, z+{\cal O}(z^2)\; ,
$$
and $\Gamma$ is the Euler function. These are two of the set of four equations for parameters of the model, to be solved in section~\ref{sec:rec}.

\subsection{Fermion propagator}
\label{subsec:fp}

To get other two equations for parameters, we put~(\ref{eq:sp}), (\ref{eq:verp}) and~(\ref{eq:dprfull}) together with~(\ref{eq:dexp}) into the DS equation~(\ref{eq:dyse}). We obtain, after some simplifications
\begin{eqnarray}
&&\!\!\!\!\!(\not\! p -m_0)S(p)=\label{eq:hsf}\\ 
&&1+[\Sigma(p)S(p)]_A+[\Sigma(p)S(p)]_B+[\Sigma(p)S(p)]_C\; ,
\nonumber
\end{eqnarray}
where we divided fermion self-energy into pieces coming from different tensor structures in $D^{\mu\nu(k)}$: $g^{\mu\nu}$ and $k^\mu k^\nu$ from the transverse part and again from the gauge-dependent longitudinal part. After small rearrangement they are:
\begin{widetext}
\begin{eqnarray}
&&\!\!\!\!\![
\Sigma (p)S(p)]_A=\label{eq:siga}\\ 
&&iZ_3\kappa^2e^2_0\int
dM\rho(M)\int\frac{d^3k}{(2\pi)^3}\left[ \gamma^{\mu}\frac{1}{\not\! p\; + \not\! k -
m+i\varepsilon}\gamma_{\mu}\frac{1}{(k^2+i\varepsilon)(k^2-\kappa^2+i\varepsilon)(\not\! p - m+i\varepsilon)} - (m\rightarrow M)\right]\nonumber\\
&&\!\!\!\!\![\Sigma (p)S(p)]_B=\label{eq:sigb}\\
&& iZ_3\kappa^2e^2_0\int dM\rho(M)\int\frac{d^3k}{(2\pi)^3}
\left[ \not\! k\frac{1}{\not\! p\; + \not\! k - m
+i\varepsilon)(k^2+i\varepsilon)^2(k^2-\kappa^2+i\varepsilon)}- (m\rightarrow M)\right]\nonumber\\
&&\!\!\!\!\![\Sigma (p)S(p)]_C=\label{eq:sigc}\\ 
&&\frac{ie^2_0}{\lambda}\int dM\rho(M)\int\frac{d^3k}{(2\pi)^3}
\left[ \not\! k\frac{1}{(\not\! p\; + \not\! k - m
+i\varepsilon)(k^2+i\varepsilon)^2} - (m\rightarrow M)\right]\nonumber\; .
\end{eqnarray}
\end{widetext}

A comment should be made here. To avoid technical difficulty while performing Wick's rotation in the above momentum integrals, we analytically continued the value of $\kappa$ to imaginary values on the upper half plane of complex $\kappa$. Now the deformation of the integration contour as is required by passing into euclidean space, is not disturbed by the inappropriate location of poles since all singularities have the `Feynman' position. After performing the integrals, we will come back to real values of $\kappa$. The procedure was discussed in~~\cite{tribb}. 
With this trick each of the above momentum integrals can be performed in an ordinary way known from perturbation theory. It can easily be seen that all integrals are finite without the need of any regularization. Omitting details of this standard calculation, we find (in Minkowski space)
\begin{widetext}
\begin{eqnarray}
I_A=&&\!\!\!\!\!ie^2_0\int\frac{d^3k}{(2\pi)^3}\gamma^{\mu}\frac{1}{\not\! p\; + \not\! k -
m+i\varepsilon}\gamma_{\mu}\frac{1}{(k^2+i\varepsilon)(k^2-\kappa^2+i\varepsilon)}=\label{eq:ia}\\
&&\!\!\!\!\!\frac{e_0^2}{8\pi\kappa^2}\int\limits_0^1dx(3m-\not\! p(1-x))\left[\frac{1}{(m^2x-p^2x(1-x))^{1/2}}-\frac{1}{(m^2x-p^2x(1-x)+\kappa^2(1-x))^{1/2}}\right]\; ,
\nonumber\\
I_B=&&\!\!\!\!\!ie^2_0\int\frac{d^3k}{(2\pi)^3}
 \not\! k\frac{1}{(\not\! p\; + \not\! k - m
+i\varepsilon)(k^2+i\varepsilon)^2(k^2-\kappa^2+i\varepsilon)}=\frac{e_0^2}{8\pi\kappa^2}\int\limits_0^1dx\bigg[\left(1-x\,\frac{\not\! p(\not\! p+m)}{\kappa^2}\right)\times\label{eq:ib}\\ &&\times\left(\frac{1}{(m^2x-p^2x(1-x))^{1/2}}
-\frac{1}{(m^2x-p^2x(1-x)+\kappa^2(1-x))^{1/2}}\right)+\frac{\not\! p(\not\! p+m)}{2}\frac{x(1-x)}{(m^2x-p^2x(1-x))^{3/2}}\bigg]\; ,
\nonumber\\
I_C=&&\!\!\!\!\! ie^2_0\int\frac{d^3k}{(2\pi)^3}
 \not\! k\frac{1}{(\not\! p\; + \not\! k - m
+i\varepsilon)(k^2+i\varepsilon)^2}=-\frac{e_0^2}{8\pi}\int\limits_0^1 dx\bigg[\frac{1}{(m^2x-p^2x(1-x))^{1/2}}\nonumber\\&&
+\frac{\not \! p(\not\! p+m)}{2}\frac{x(1-x)}{(m^2x-p^2x(1-x))^{3/2}}\bigg]\; .
\label{eq:ic}
\end{eqnarray}
\end{widetext}

Now the contributions to $\Sigma(p)S(p)$ may be written as
\begin{eqnarray}
&&\!\!\!\!\! [\Sigma (p)S(p)]_A =\label{eq:ssa}\\
&&Z_3\kappa^2\int dM\rho(M)\left[I_A\frac{1}{\not\! p-m+i\varepsilon}-(m\rightarrow M)\right]\; ,\nonumber\\
&&\!\!\!\!\! [\Sigma (p)S(p)]_B =\label{eq:ssb}\\
&&Z_3\kappa^2\int dM\rho(M)\left[I_B-(m\rightarrow M)\right]\; ,\nonumber\\
&&\!\!\!\!\! [\Sigma (p)S(p)]_C =\frac{1}{\lambda}\int dM\rho(M)\left[I_C-(m\rightarrow M)\right]\label{eq:ssc}\; .
\end{eqnarray}
All the above spectral integrals can been found similarly as in~\cite{tr3}, therefore we omit the technicalities. 

In the infrared domain, when $p^2\rightarrow m^2$, the left hand side of the DS equation~(\ref{eq:hsf}), after substituting~(\ref{eq:spr}), contains two types of singular terms (the only terms that are important):
\begin{equation}
-\frac{m^{2\beta}\delta m (\not\! p +m)}{(m^2-p^2)^{\beta+1}}+\frac{m^{2\beta}}{(m^2-p^2)^{\beta}}\; .
\label{eq:singl}
\end{equation}
Therefore, to avoid lengthy expressions we will not give the results for~(\ref{eq:ssa}), (\ref{eq:ssb}), and~(\ref{eq:ssc}) in their full complexity, because it is sufficient for our goal to only pick out from them the identical singular terms. We find:
\begin{eqnarray}
&&\!\!\!\!\! [\Sigma (p)S(p)]_A \approx \frac{e_0^2Z_3}{8\pi}\bigg\{\bigg[\frac{1}{m^2}\,{\cal I}_1(m^2,\kappa^2)-\frac{1}{m^3}\bigg]\times\nonumber\\
&&\times\frac{\not\! p(\not\! p+m)}{(1-p^2/m^2)^{\beta+1}}+\bigg[\frac{\not\! p(\not\! p+m)}{2}\left({\cal I}_2(m^2,\kappa^2)-\frac{1}{m^3}\right)\nonumber\\
&&-\frac{3}{4\beta m}\bigg]\frac{1}{(1-p^2/m^2)^{\beta}}\bigg\}\; ,\label{eq:ssa1}
\end{eqnarray}
\begin{eqnarray}
&&\!\!\!\!\! [\Sigma (p)S(p)]_B \approx \frac{e_0^2Z_3}{8\pi}\bigg[\frac{1}{m^3}\,\frac{\not\! p(\not\! p+m)}{(1-p^2/m^2)^{\beta+1}}\label{eq:ssb1}\\
&&-\bigg(\frac{\not\! p(\not\! p+m)}{2m^3}\, \frac{1-\beta}{\beta}-\frac{1}{4\beta m}\bigg)\frac{1}{(1-p^2/m^2)^{\beta}}\bigg]\; ,\nonumber
\end{eqnarray}
\begin{eqnarray}
&&\!\!\!\!\! [\Sigma (p)S(p)]_C \approx -\frac{e_0^2}{8\pi\lambda}\bigg[\frac{\not\! p(\not\! p+m)}{m^3}\,\frac{1}{(1-p^2/m^2)^{\beta+1}}\nonumber\\
&&+\bigg(\frac{\not\! p(\not\! p+m)}{2m^3}\,\frac{\beta-1}{\beta}+\frac{1}{4\beta m}\bigg)\frac{1}{(1-p^2/m^2)^{\beta}}\bigg]\label{eq:ssc1}\; ,
\end{eqnarray}
where $\approx$ refers to diverging terms, when $p^2\rightarrow m^2$. The functions ${\cal I}_1$ and ${\cal I}_2$ have the following form:
\begin{eqnarray}
{\cal I}_1(m^2,\kappa^2)&\!\!\! =&\!\!\!\int\limits_0^1 dx\frac{x+2}{(m^2x^2+\kappa^2(1-x))^{1/2}}\; ,
\label{eq:ii1}\\
{\cal I}_2(m^2,\kappa^2)&\!\!\! =&\!\!\!\int\limits_0^1 dx\frac{x^2(x+2)}{(m^2x^2+\kappa^2(1-x))^{3/2}}\; .
\label{eq:ii2}
\end{eqnarray}

If the solution is to be self-consistent close to the fermion mass shell, the divergent terms on both sides of DS equation~(\ref{eq:hs}) must be identical. Equating them, and making use of the fact that up to finite terms we have
\begin{eqnarray}
\frac{\not\! p(\not\! p+m)}{(1-p^2/m^2)^{\beta}}&\!\! \approx &\!\!\frac{2m^2}{(1-p^2/m^2)^{\beta}}\; ,\label{eq:apm}\\
\frac{\not\! p(\not\! p+m)}{(1-p^2/m^2)^{\beta+1}}&\!\!\approx &\!\!\frac{m(\not\! p+m)}{(1-p^2/m^2)^{\beta+1}}-\frac{m^2}{(1-p^2/m^2)^{\beta}}\; ,\nonumber
\label{eq:aapm}
\end{eqnarray}
we derive the following two relations for unknown parameters $\delta m$ and $\beta$:
\begin{eqnarray}
\delta m &\!\!\! =&\!\!\! \frac{e_0^2 Z_3}{8\pi}\left[-m{\cal I}_1(m^2,\kappa^2)+\frac{1}{\lambda Z_3}\right]\; ,\label{eq:eq3}\\
1 &\!\!\! =&\!\!\! \frac{e_0^2 Z_3}{8\pi}\left[-\kappa^2{\cal I}_3(m^2,\kappa^2)+\frac{3}{4\beta m}\left(\frac{1}{\lambda Z_3}-2\right)\right]\; .\nonumber\\ \label{eq:eq4}
\end{eqnarray}
The function ${\cal I}_3(m^2,\kappa^2)$ is defined as follows
\begin{eqnarray}
{\cal I}_3 =&&\!\!\!\!\frac{1}{\kappa^2}\left({\cal I}_1(m^2,\kappa^2)-m^2{\cal I}_2(m^2,\kappa^2)\right]\nonumber=\\
&&\!\!\!\!\int\limits_0^1 dx\frac{(1-x)(x+2)}{(m^2x^2+\kappa^2(1-x))^{3/2}}\; .
\label{eq:ii3}
\end{eqnarray}

\section{Solutions and conclusions}
\label{sec:rec}

The four equations we have obtained, i.e~(\ref{eq:eq1}), (\ref{eq:eq2}), (\ref{eq:eq3}) and~(\ref{eq:eq4}) are sufficient to determine all parameters. We rewrite them with the use of {\em renormalized} quantities: fermion mass $m=m_0+\delta m$, gauge coupling constant $e=Z_3^{1/2} e_0$ and gauge parameter $\lambda_R=Z_3\lambda$. Besides, it is useful to introduce a dimensionless parameter $\zeta=\frac{e^2}{4\pi m}$. After executing the parametric integrals in ${\cal I}_1$ and ${\cal I}_3$ and performing the reverse analytical continuation in $\kappa$, we get

\begin{figure}[b]
\centering
{\includegraphics[width=0.45\textwidth]{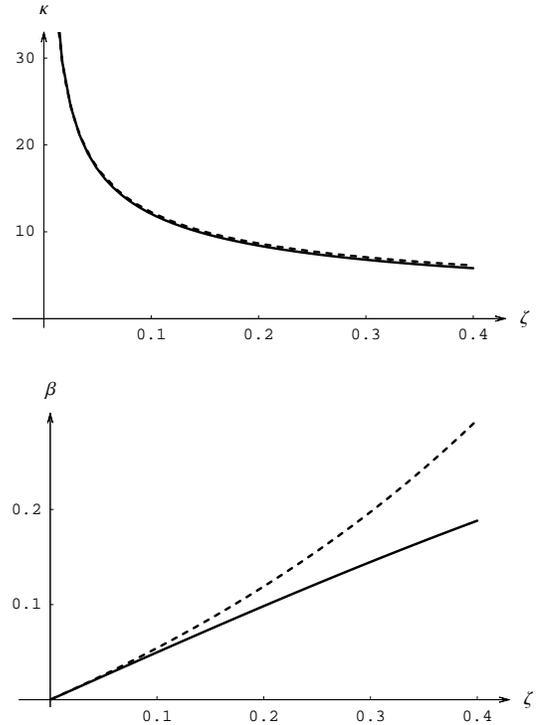}
\caption{The dependence of $\kappa$ in units of $m$ (upper plot) and of power $\beta$ (lower plot) on the parameter $\zeta$. The dashed line corresponds to approximated solutions defined by equations~(\ref{eq:deleqa})-(\ref{eq:kapeqa}). The gauge parameter is chosen as $\lambda_R=0.3$.} \label{fig:graph1}}
\end{figure} 

\begin{eqnarray}
\frac{\delta m}{m} = &&\!\!\!\!\!\frac{\zeta}{2}\bigg[\frac{1}{\lambda_R}-1-\frac{4m^2-\kappa^2}{4m^2}\,\ln (4m^2/\kappa^2+1)\bigg],\label{eq:deleq}\\
1 = &&\!\!\!\!\!\frac{\zeta}{2}\bigg[\frac{3}{4\beta }\left(\frac{1}{\lambda_R}-2\right)-2\,\frac{4m^2-\kappa^2}{4m^2+\kappa^2}\nonumber\\
&&\!\!\!\!\!-\frac{\kappa^2}{2m^2}\,\ln (4m^2/\kappa^2+1)\bigg]\; ,\label{eq:jedeq}\\
Z_3 = &&\!\!\!\!\! 1-\frac{2\zeta}{3}\frac{\Gamma(\beta+1/2)}{\sqrt{\pi}\Gamma(\beta+1)}\; ,\label{eq:z3eq}\\
\frac{\kappa^2}{m^2} = &&\!\!\!\!\! \frac{15  }{2\zeta}\frac{\sqrt{\pi}\Gamma(\beta+1)}{\Gamma(\beta+3/2)}\; .
\label{eq:kapeq}
\end{eqnarray} 

\begin{figure}[b]
\centering
{\includegraphics[width=0.45\textwidth]{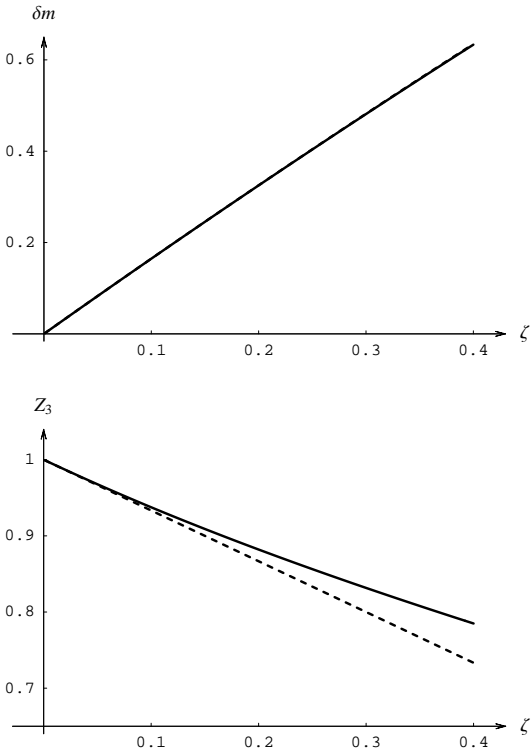}
\caption{The dependence of mass renormalization $\delta m$ in units of $m$ (upper plot) and charge renormalization constant $Z_3$ (lower plot) on the parameter $\zeta$.  The dashed line corresponds to approximated solutions defined by equations~(\ref{eq:deleqa})-(\ref{eq:kapeqa}). On the upper plot the dashed line is almost identical with the solid one and therefore it is not visible. The gauge parameter is chosen as $\lambda_R=0.3$.} \label{fig:graph2}}
\end{figure}

This set of equations for $\delta m/m$, $\beta$, $Z_3$ and $\kappa^2/m^2$ may be solved numerically for certain chosen values of renormalized gauge parameter $\lambda_R$ and the results plot as functions of parameter $\zeta$.
Let us consider the case of weak coupling, when $\zeta$ is small. Since the applicability of our method requires $0<\beta<1$, then from equation~(\ref{eq:kapeq}) we deduce that $\kappa^2/m^2$ should be large. 
It may be justified by the elementary estimation given below. First we rewrite following expression in terms of beta function $B(x,y)$ and its integral representation:
\begin{equation}
\frac{\sqrt{\pi}\Gamma(\beta+1)}{2\Gamma(\beta+3/2)}=\frac{1}{2}\, B(\beta+1, 1/2)=\frac{1}{2}\int\limits_0^1 t^{-1/2}(1-t)^\beta dt\; ,\nonumber
\end{equation}
and next use the inequalities valid for $0<\beta<1$:
\begin{eqnarray}
\frac{1}{2}\int\limits_0^1 t^{-1/2}(1-t)^\beta dt&\! <&\! \frac{1}{2}\int\limits_0^1 t^{-1/2} dt=1\; ,\nonumber\\
\frac{1}{2}\int\limits_0^1 t^{-1/2}(1-t)^\beta dt&\! >&\! \frac{1}{2}\int\limits_0^1 t^{-1/2}(1-t) dt=\frac{2}{3}\; .\nonumber
\end{eqnarray}
Now, from~(\ref{eq:kapeq}) it becomes obvious that for small $\zeta$ the left hand side must be large. This is in agreement with our expectations concerning the subsequent terms in the Taylor expansion~(\ref{eq:dexp}). But for $\kappa^2\gg m^2$ the first term in square brackets on the right hand side of~(\ref{eq:jedeq}) dominates over all other, which may be easily verified. The expected positivity of $\beta$ requires then considering only gauges for which $0<\lambda_R<1/2$. This is not surprising, since $\beta$ is a strongly gauge dependent quantity. For our numerical calculations we have then chosen the values of the gauge parameters from that range.

In figure~\ref{fig:graph1} we show the dependence of parameters $\kappa$ and $\beta$ on $\zeta$ for exemplary value of $\lambda_R=0.3$ (solid lines). In figure~\ref{fig:graph2} we similarly plot the dependence of two other parameters: fermion mass renormalization constant $\delta m$ and gauge field renormalization constant $Z_3$. It is nice to observe that we have $0<\delta m <m$ and $0<Z_3<1$, as we expected. The dashed lines in these figures represent approximate solutions of the set (\ref{eq:deleq})-(\ref{eq:kapeq}), as described below.

\begin{figure}[b]
\centering
{\includegraphics[width=0.45\textwidth]{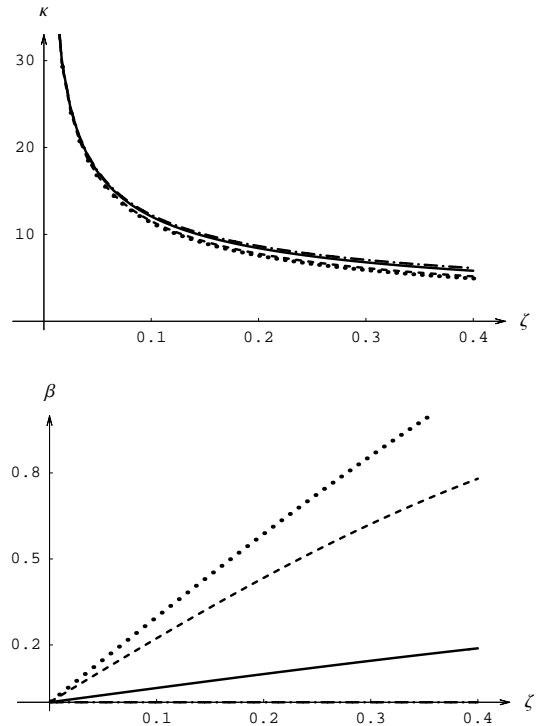}
\caption{The comparison of the behavior  of $\kappa$ (upper plot) and parameter $\beta$ (lower plot) for different values of gauge parameter: dotted line -- $\lambda_R=0.1$, dashed line -- $\lambda_R=0.2$, solid line -- $\lambda_R=0.3$, mixed line -- $\lambda_R=0.5$ (Yennie gauge). The plots of $\kappa$ for various gauges follow almost the same curve.} \label{fig:graph3}}
\end{figure}

In the weak coupling regime, assuming that we do not choose the gauge parameter $\lambda_R$ approaching zero (but still to be less than $1/2$), the equations may be given the following approximated form:
\begin{eqnarray}
\frac{\delta m}{m} &\!\!\! =&\!\!\! \frac{\zeta}{2}\left[\frac{1}{\lambda_R}-\frac{6m^2}{\kappa^2}\right]\; ,\label{eq:deleqa}\\
1&\!\!\! =&\!\!\! \frac{\zeta}{2}\left[\frac{3}{4\beta }\left(\frac{1}{\lambda_R}-2\right)-\frac{12m^2}{\kappa^2}\right]\; ,\label{eq:jedeqa}\\
Z_3&\!\!\! =&\!\!\! 1-\frac{2\zeta}{3}\; ,\label{eq:z3eqa}\\
\frac{\kappa^2}{m^2}&\!\!\! =&\!\!\! \frac{15 }{\zeta}\; .
\label{eq:kapeqa}
\end{eqnarray}
The terms $m^2/\kappa^2$ in the first two equations arise from the expansion of ${\cal I}_1$ and ${\cal I}_3$ for large $\kappa^2$, but in fact they may be omitted, since they are of order $\zeta$, as results from the last equation. As already told, the solutions of these simplified equations are drawn in figures~\ref{fig:graph1} and~\ref{fig:graph2} as dashed lines. On the plots for $\kappa$ and $\delta m$ these curves are not visible since they are almost identical with the full solutions, but also on the other two they do not deviate from the `exact' results too much.

\begin{figure}[t]
\centering
{\includegraphics[width=0.45\textwidth]{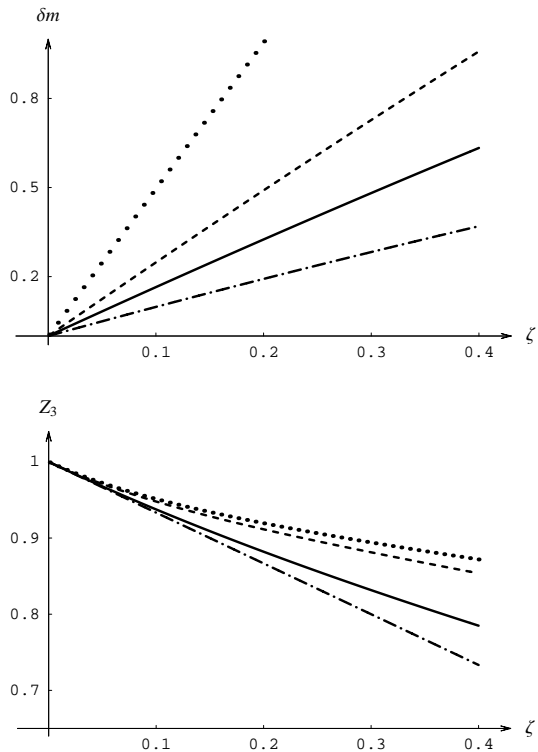}
\caption{The comparison of the behavior of fermion mass renormalization $\delta m$ (upper plot) and gauge field renormalization constant $Z_3$ (lower plot) for different vales of gauge parameter: dotted line -- $\lambda_R=0.1$, dashed line -- $\lambda_R=0.2$, solid line -- $\lambda_R=0.3$, mixed line -- $\lambda_R=0.5$ (Yennie gauge).} \label{fig:graph4}}
\end{figure}

From~(\ref{eq:jedeqa}) we see that the exponent $\beta$ in this approximation may be written as
\begin{equation}
\beta=\frac{3\zeta}{8}\left(\frac{1}{\lambda_R}-2\right)\; ,
\label{eq:beje}
\end{equation}\\
\\
which means that $\lambda_R=1/2$ is a kind of Yennie gauge.

It is interesting to observe, if and how the values of the parameters depend on the gauge. This is shown in figures~\ref{fig:graph3} and~\ref{fig:graph4} which are performed for the following values from the range $[0,1/2]$: $\lambda_R=0.1,\; 0.2,\; 0.3,\; 0.5$. The Landau gauge cannot be used because it would lead to negative value of $\beta$ .

The particular stress deserves the observation that the gauge dependence of the parameter $\kappa$ is extremely weak. This means that the renormalized gauge boson propagator (strictly speaking its transverse part) is practically gauge independent, as it should be. Please note that in other approximation schemes applied to DS equations in QED$_3$ one obtains the gauge dependent value of the polarization scalar~\cite{bpr}.

The dependence of $Z_3$ on $\lambda_R$ is relatively weak for small $\zeta$ too. These results are worth noticing, since the full gauge independence should appear in the {\em exact} theory, and one ought not to expect too much from the approximated model, where infrared forms of Green's functions are postulated in a simple form. Gauge dependence of the `physical' fermion mass obtained in our work is relatively strong but it is a common feature of nonperturbative calculations in this model~\cite{bas1,bas2,bara2}.

\end{document}